\documentclass[12pt]{article}
\usepackage[a4paper,left=3.0cm,right=3.0cm,top=3.0cm,bottom=3.0cm]{geometry} 
\usepackage[latin1]{inputenc} 
\usepackage{setspace} 
\usepackage{fancyhdr} 
\usepackage{amsmath} 
\usepackage{amssymb} 
\usepackage{amsfonts} 
\usepackage{dsfont} 
\usepackage{amsthm} 
\usepackage{graphicx} 
\usepackage{tikz} 
\usepackage{color} 
\definecolor{goetheblue}{RGB}{0,102,255} 
\usepackage{listings} 
\usepackage{eurosym} 
\usepackage{url} 
\usepackage{enumerate} 
\usepackage{multirow} 
\usepackage{verbatim} 
\usepackage{mathrsfs}
\usepackage{booktabs}
\usepackage{multirow}
\usepackage{adjustbox}		
\usepackage[round]{natbib}
\usepackage{url}
\usepackage{appendix}

\newcommand{\p}{\mathrm{P}}
\newcommand{\E}{\mathrm{E}}

\begin{document}

\title{Unlucky Number 13?\\ {M}anipulating {E}vidence {S}ubject to {S}nooping\thanks{\textbf{Acknowledgements:} We are grateful to Matei Demetrescu, Steffen Eibelshäuser, Mehdi Hosseinkouchack, Michael Neugart and Jan Reitz for many helpful comments.}}
		
\author{Uwe Hassler\thanks{Corresponding Author: E-Mail: hassler@wiwi.uni-frankfurt.de, Address: Goethe University Frankfurt, RuW Building, Theodor-W.-Adorno-Platz 4, 60323 Frankfurt, Germany} and Marc-Oliver Pohle\\	Goethe University Frankfurt} 

\maketitle

\begin{abstract}
	
Questionable research practices like HARKing or $p$-hacking have generated considerable recent interest throughout and beyond the scientific community. We subsume such practices involving secret data snooping that influences subsequent statistical inference under the term MESSing (manipulating evidence subject to snooping) and discuss, illustrate and quantify the possibly dramatic effects of several forms of MESSing using an empirical and a simple theoretical example. The empirical example uses numbers from the most popular German lottery, which seem to suggest that 13 is an unlucky number. 
	
\end{abstract}

\newline

\section{Introduction}

\begin{figure}
	\noindent
	\centering{}\includegraphics[scale=0.6]{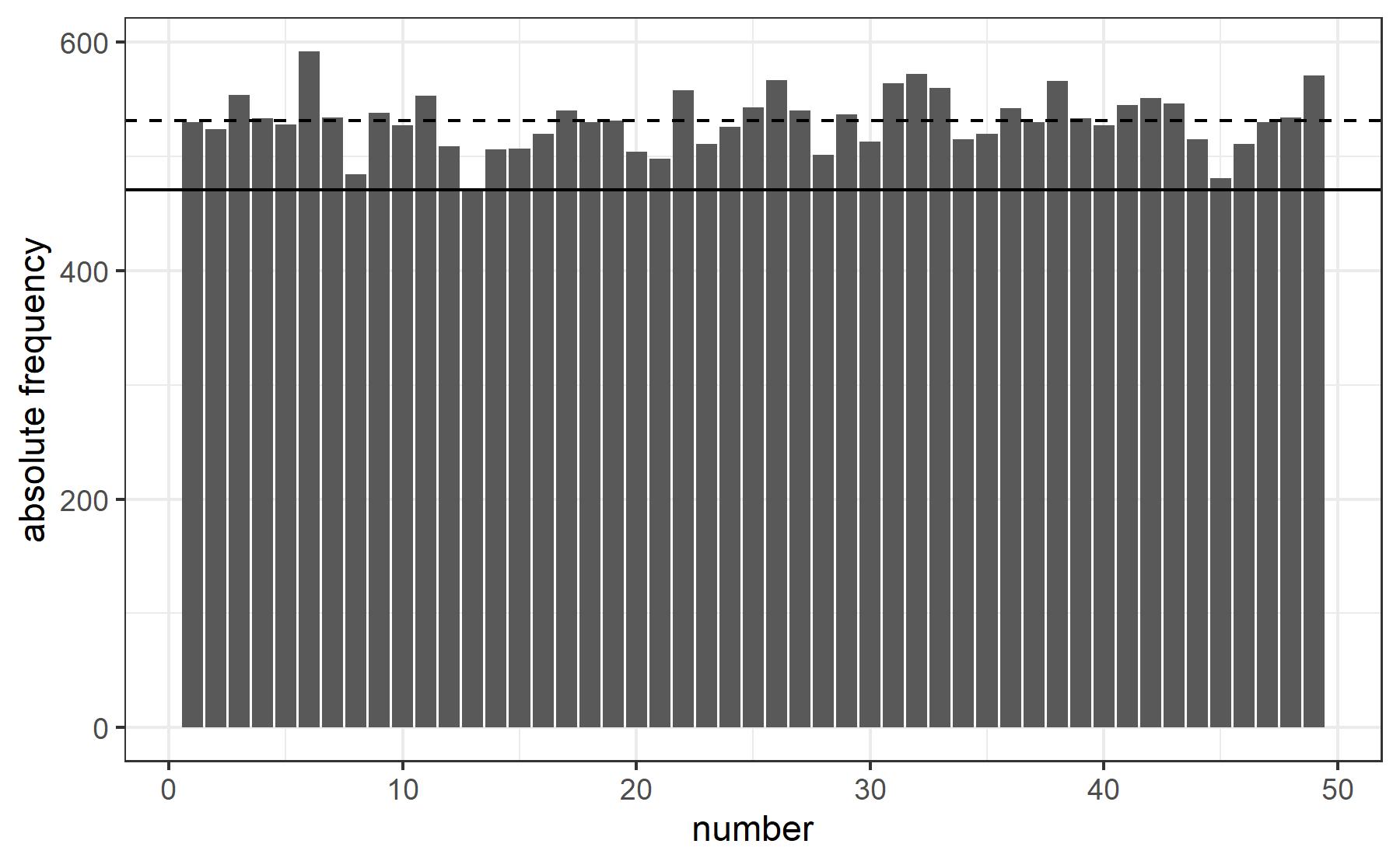}
	\caption{Frequency distribution of $N=4337$ games, i.e.\ from $n=6\cdot N =26022$ numbers (German Lotto 6 out of 49); broken line: expected frequency under equal probability, solid line: frequency of number 13} 
	\label{fig1}	
\end{figure}

The most popular lottery in Germany is ``Lotto 6 out of 49''. The 6 winning numbers of one game are determined by drawing 6 balls without replacement from a pool of 49 balls identified by means of  the first 49 natural numbers. The first game took place on October 9, 1955, and the first ball ever drawn carried the number 13.
Until November 29, 2019, a total of 4337 games had been played with $n= 6 \cdot 4337 =26022$ balls being drawn. Figure \ref{fig1} displays the absolute frequencies for each number 1 through 49.\footnote{The data were downloaded from https://www.lotto.de/lotto-6aus49/statistik/ziehungshaeu\-figkeit on November 29, 2019.}
What does strike you at first glance? The number 13 stands out with the least favourable odds.  This may come as no surprise to people that consider 13 to be an ``unlucky number''. As a result of fear of the number 13 (clinically: triskaidekaphobia) there is no row 13 in many planes or no floor 13 in many tall buildings and many people avoid Friday the 13th for marriage. And indeed,  the number 13 was drawn only 471 times in the German lottery, while (roughly) 531 cases would have to be expected under equal probability of all 49 numbers given 26022 draws (broken line in Figure \ref{fig1}). If a PhD student presents such descriptive evidence to her or his supervisor, the supervisor might ask: Is the deviation significant? At which level? And how can we test properly?

Let $p_{13}$ be the probability of getting 13 when drawing a Lotto number. We wish to test  $p_{13} = 1/49$ against the one-sided alternative:
\[
H_0: \, p_{13} = \frac{1}{49} \quad \mbox{vs.} \quad H_1: \, p_{13} < \frac{1}{49} \, .
\] The test statistic accounting for the implied dependence of the German lottery due to drawing without replacement is  $Z_{13}^{(6)}$ from equation (\ref{Zmod}) resulting in $Z_{13}^{(6)}=-2.7822$ with the highly significant  $p$-value of 0.270\% relying on the normal approximation. Is the German lottery flawed? Is 13 truely an unlucky number? What is going wrong?

The answer to the last question is what we call MESSing. We propose the generic term MESSing (Manipulating Evidence Subject to Snooping) to subsume practices which involve conducting statistical inference after data analysis of some form has already been carried out and influenced the researchers' decisions, but is not acknowledged. 

In recent years a number of such questionable research practices related to statistical inference situated somewhere in the grey zone between exploratory data analysis and fraud have attracted the attention of the academic literature in many fields such as medicine, psychology or economics (see e.g.\ \cite{ioannidis2005}, \cite{Simmonsetal2011} and \cite{brodeur2016}). Popular examples are HARKing (Hypothesizing After the Results are Known) (see \cite{Kerr1998}) or $p$-hacking (see \cite{Simmonsetal2011}). Often these practices seem innocent, are not used with bad intentions and are deeply rooted in the research culture. But they invalidate inference and may consequently lead to wrong results and distorted literatures. Further, due to the lack of transparency related to these practices their extent and exact consequences are very hard to assess. Even though their detrimental effects like e.g.\ impeding the replicability of research, slowing down scientific progress or damaging the credibility of science have been discussed and some steps have been taken to curb them in recent years (see e.g.\ \cite{Wasserstein2019}
 or \cite{christensen2019}), they still seem to be in widespread use. Convincing researchers to refrain from these practices is, alongside other measures, certainly crucial. This starts with a clear understanding of the issue and the problems arising from it throughout the scientific community. Our paper aims to contribute to that. We give an overarching definition subsuming all these practices under the term MESSing and discuss and illustrate several forms of MESSing using the example of the German lottery numbers and a simple theoretical example. While some forms of MESSing like HARKing or $p$-hacking are well documented, we also point out other forms that have hardly or not at all been recognized. These include practices trying to weaken evidence instead of strengthening it and driving evidence to the extreme instead of just jumping over significance thresholds. 


The next section explains why the ``puzzle'' of the overly significant unlucky number 13 arises and how it can be solved. Then several forms of MESSing are illustrated using the example of the lottery numbers. Section 3 connects this discussion to previous literature  and goes into some details on MESSing. Section 4 contains a simple theoretical example that illustrates and allows to quantify the effects of MESSing in different directions. Some conclusions are offered in the final section.

\section{Testing for Unlucky 13}

The general case of Lotto $K$ out of $V$ consists of $K$ balls drawn without replacement in one game from an urn of $V$ balls. In the German lottery, we have $V=49$ and $K=6$, but we stick with a general $K$ for reasons that will become obvious below.

Let $L_1, \cdots, L_K, L_{K+1}, \cdots,  L_{2 \cdot K}, L_{2 \cdot K +1}, \cdots,  L_{N \cdot K}$ be the consecutive numbers drawn in $N$ games. To execute a test of uniformity we are interested in the counts of the 49 numbers from the sample of size $n = N \cdot K$. Let the counts of these numbers be denoted by $S_m, m=1,2,...,49$, and consider the Bernoulli random variables
$$X_{m,i}=\begin{cases} 1 &\text{if } L_i=m\\
0 &\text{if } L_i \neq m \end{cases}, \ i= 1 \ldots, N \cdot K = n,$$ which indicate if the $i$th ball drawn shows the number $m$ or not. These are the ingredients to determine the total counts 
$$S_m = \sum_{i=1}^n X_{m,i}.$$

We are interested in testing the null hypothesis
$$H_0: P(X_{m,i}=1) = \frac 1 {49} \text{ for  one  specific } m \in \{1,2,...,49\}.$$
Under $H_0$ it holds that 
$X_{m,i} \sim Be \left( \frac 1 {49} \right)$ 
and
$\E[S_{m}] = \frac n {49}$. However, due to the dependence between the Bernoulli random variables $X_{m,i}$ within one game caused by drawing without replacement,  
$S_{m}$ does not follow a binomal distribution $Bi \left(n,\frac 1 {49} \right)$. Consequently, the classical binomial test constructed under the i.i.d. assumption,
\begin{equation} \label{Ziid}
Z_{m}^{iid} := \frac{S_{m} - \frac n {49}}{\sigma_{iid}} \text{ with } \sigma_{iid}^2 :=\frac{n(49-1)}{49^2},
\end{equation}
does not follow a standard normal law asymptotically. However, as we show in the appendix, only the variance decreases due to the negative dependence between the $X_{m,i}$,
\begin{equation} \label{variances}
\sigma_{(K)}^2 := \frac{49-K}{49-1} \sigma_{iid}^2,
\end{equation} 
and has to be adjusted to retain limiting standard normality under $H_0$:
\begin{equation} \label{Zmod}
Z_{m}^{(K)} :=\frac{S_{m} - \frac{n}{49}}{\sigma_{(K)}}=\sqrt{\frac{49-1}{49-K}}Z_{m}^{iid} \ \xrightarrow[]{d} \  \mathcal{Z},
\end{equation}
where $\mathcal{Z}$ follows a standard normal distribution, $\mathcal{Z} \sim \mathcal{N}(0,1)$, and ``$\xrightarrow[]{d}$'' denotes convergence in distribution as the sample size $n$ diverges.

As mentioned in the introduction, when computing this test statistic for $m=13$, we get $Z_{13}^{(6)}=-2.7822$ with the highly significant $p$-value of 0.00270. What causes this test result is of course MESSing. This nonsensical significance arises because we first looked at the data in Figure \ref{fig1}, observed the remarkable deviation of $S_{13} = 471$ from $\frac {26022} {49} = 531.06$, and then tested for this specific hypothesis. A real MESSy would tell a more or less convincing story here or present some theory that lead him or her to come up with this hypothesis, which would in our case of the unlucky number 13 be quite comical. In other cases, this is usually not that obvious and MESSing is hard to detect. 

Such a MESS had been blamed already by \citet[p. 229]{Wallis1942}: ``An investigator who after inspecting the data decides what to test or how to make the test can, by virtue of the fact that any sample has unique characteristics, disprove any hypothesis.'' This form of MESSing, where the same data is used to postulate and to test the hypothesis, has later been called HARKing, which will be discussed in detail in the next section. 

\begin{figure}
	\centering
	\includegraphics[width=0.7\linewidth]{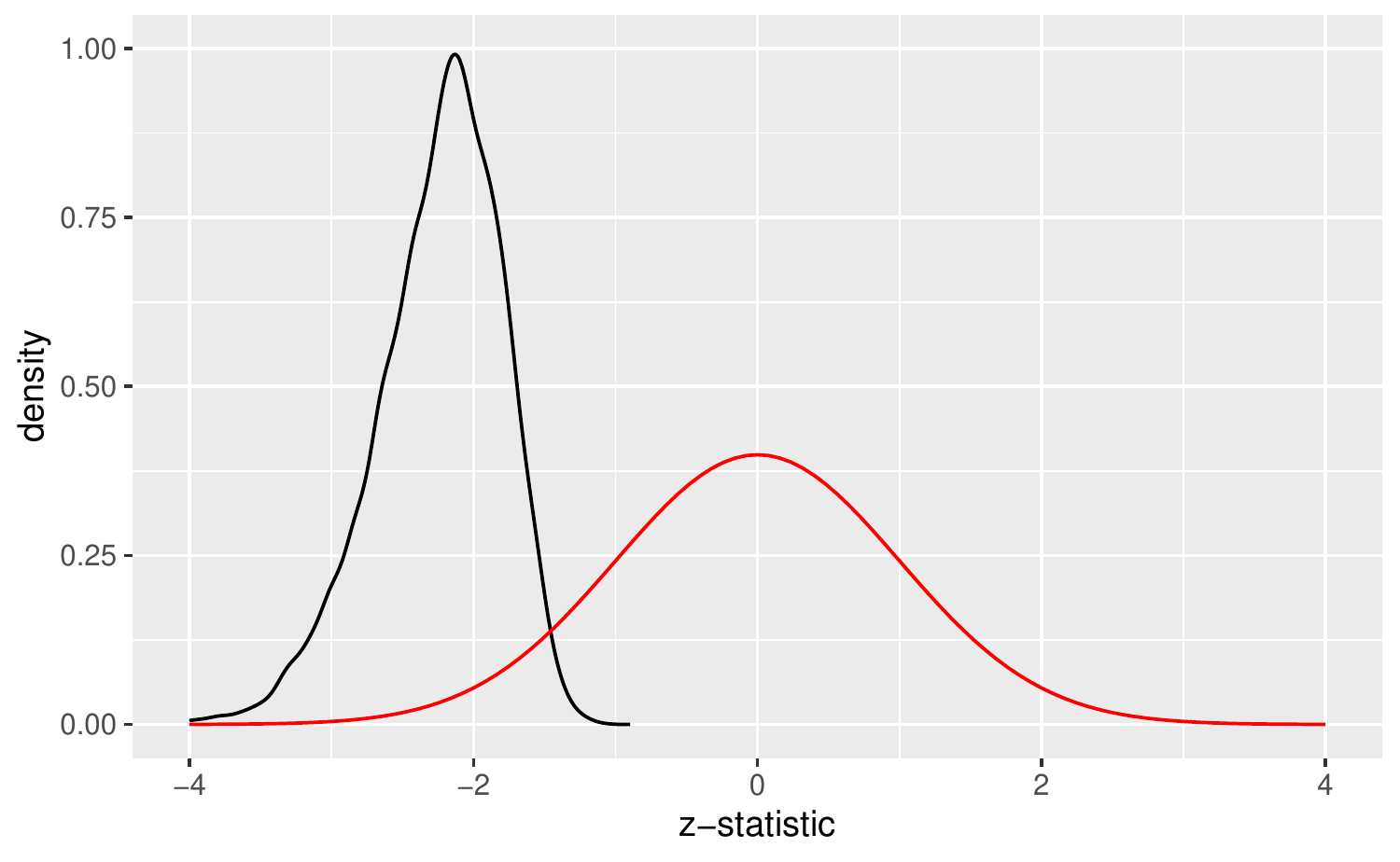}
	\caption{Density of $\mathcal{Z}_{\min}^{(6)}$ (obtained by simulations)  compared to a standard normal density}
	\label{fig:zstats}
\end{figure}

We are able to quantify the effect of this MESS in our example if we are willing to assume that the Lotto numbers follow a uniform distribution: What we did amounts to testing with  $\min_{m=1, \ldots, 49} S_m$, and not surprisingly, the minimum deviates significantly from the overall mean. The number $13$ was picked for testing because $m=13$ leads to the strongest left-sided violation of the null in favour of $p_{m} < 1/49$. Let $\min S := \min_{m=1, \ldots, 49} S_m$ and
\[
Z_{min}^{(6)} := \frac{\min S - \frac{n}{49}}{\sigma_{(6)}} .
\]
We denote the limit in distribution of $Z_{min}^{(6)}$ for $n \rightarrow \infty$ as $\mathcal{Z}_{min}^{(6)}$. The density of $\mathcal{Z}_{min}^{(6)}$ is depicted in Figure \ref{fig:zstats} alongside a standard normal density, i.e.\ the density of the asymptotic distribution of $Z_{m}^{(6)}$ under the null.\footnote{We approximated the distribution of $\mathcal{Z}_{min}^{(6)}$ under uniformity of Lotto numbers by simulations, i.e.\ we simulated 10000 times a sample of $6 \cdot 10000$ Lotto numbers and then looked at the empirical distribution of the test statistics. The density, which is drawn in the picture, is a kernel density estimate. Of course, we could also have simulated for our sample size $N=4337$, but there are virtually no differences as we are very close to the asymptotic distribution no matter if $N=4337$ or 10000.} Due to MESSing the distribution dramatically changes its shape, leading to very small $p$-values. 
From $\mathcal{Z}_{min}^{(6)}$ we can also calculate the size distortions: The rejection probabilities under the null hypothesis, i.e.\ the type I errors, for the three significance levels 0.01, 0.05 and 0.1 are (with $z_{\alpha}$ denoting the $\alpha$-quantile of $\mathcal{Z} \sim \mathcal{N}(0,1)$):
\[
P(\mathcal{Z}_{min}^{(6)} \leq z_{0.01}) = 0.3879, P(\mathcal{Z}_{min}^{(6)} \leq z_{0.05}) = 0.9376 \text{ and } P(\mathcal{Z}_{min}^{(6)} \leq z_{0.1}) = 0.9984.
\]

In our empirical example with $Z_{13}^{(6)}=-2.7822$ the $p$-value corrected for MESSing is $P(\mathcal{Z}_{min}^{(6)} \leq -2.7822) = 0.1203$. However, in practice, such a correction for MESSing is usually infeasible as there are many forms of MESSing and it is usually unknown which forms have been used if any. 

There may be statistical methods that are more robust to certain forms of MESSing. In our case, such a method is available: A proper way to test the null hypothesis of uniformity has to take into account \emph{each} number $m \in \{1, \ldots, 49\}$, which amounts to the  goodness-of-fit test by \citet{Pearson1900} for the joint  hypothesis:
\begin{equation} \label{jointH}
H_0: P(X_{1,i}=1)=P(X_{2,i}=1)=...=P(X_{49,i}=1)=\frac 1 {49}.
\end{equation}
Again the dependence in the $X_{m,i}$ due to drawing without replacement invalidates the classical approach, i.e.\ the test statistic
\begin{equation} \label{chi_iid}
\chi^2_{iid} := \sum_{m=1}^{49} \frac{( S_m - \frac{n}{49})^2}{\frac{n}{49}}
\end{equation}
does not converge in distribution to a $\chi^2(48)$ random variable. Again, only a scaling factor (known from equation (\ref{variances})) is required to recover the limiting chi-squared distribution under the null:
\begin{equation} \label{chi_mod_K}
\chi^2_{(K)} := \frac{49-1}{49-K}\,\chi^2_{iid} \ \stackrel{d}{\to} \ \chi^2(48). 
\end{equation}
This limit arises as a special case of more general results by \citet[p. 183]{Joe1993} for Pearson's tests for uniformity of $k$-tuples, $k \in \{1,2,...,K\}$ in the lottery $K$ out of $V$; see  also \cite{Genestetal2002} or the earlier closely related results from the survey sampling literature by \cite{RaoScott1981}. The data behind Figure \ref{fig1} provide $\chi^2_{iid}= 55.10$ and $\chi^2_{(6)}= 61.51$, see Table \ref{tab_lotto}. 
The $p$-value of $\chi^2_{(6)}$ when comparing with the $\chi^2 (48)$ distribution is 9.11\%. The $p$-value is much higher now and the German lottery data do not violate the uniform distribution hypothesis (\ref{jointH}) at a 9\% significance level.

\begin{table}[]
	\centering
	\begin{tabular}{@{}lllll@{}}
		\toprule
		sample & type &  $n$ & statistic & $p$-value\\ \midrule
		I & 6/49 & 26022   & $\chi^2_{(6)} = 61.51$ & 9.11\% \\ 
		II & 6/49 + add. num. & 29637  & $\chi^2_{(7)} = 60.58$ & 10.51\%  \\
		III & add. num. only & 3615 & $\chi^2_{iid} = 48.64$ & 44.70\%  \\ \bottomrule
	\end{tabular}
	\caption{Sample I: numbers ``6 out of 49'' without replacement from 4337 games; sample II: sample I plus an additional number drawn without replacement in 83.35\% of all games; sample III: independently drawn additional number only}
	\label{tab_lotto}
\end{table}

\begin{figure}
	\noindent
	\centering{}\includegraphics[scale=0.6]{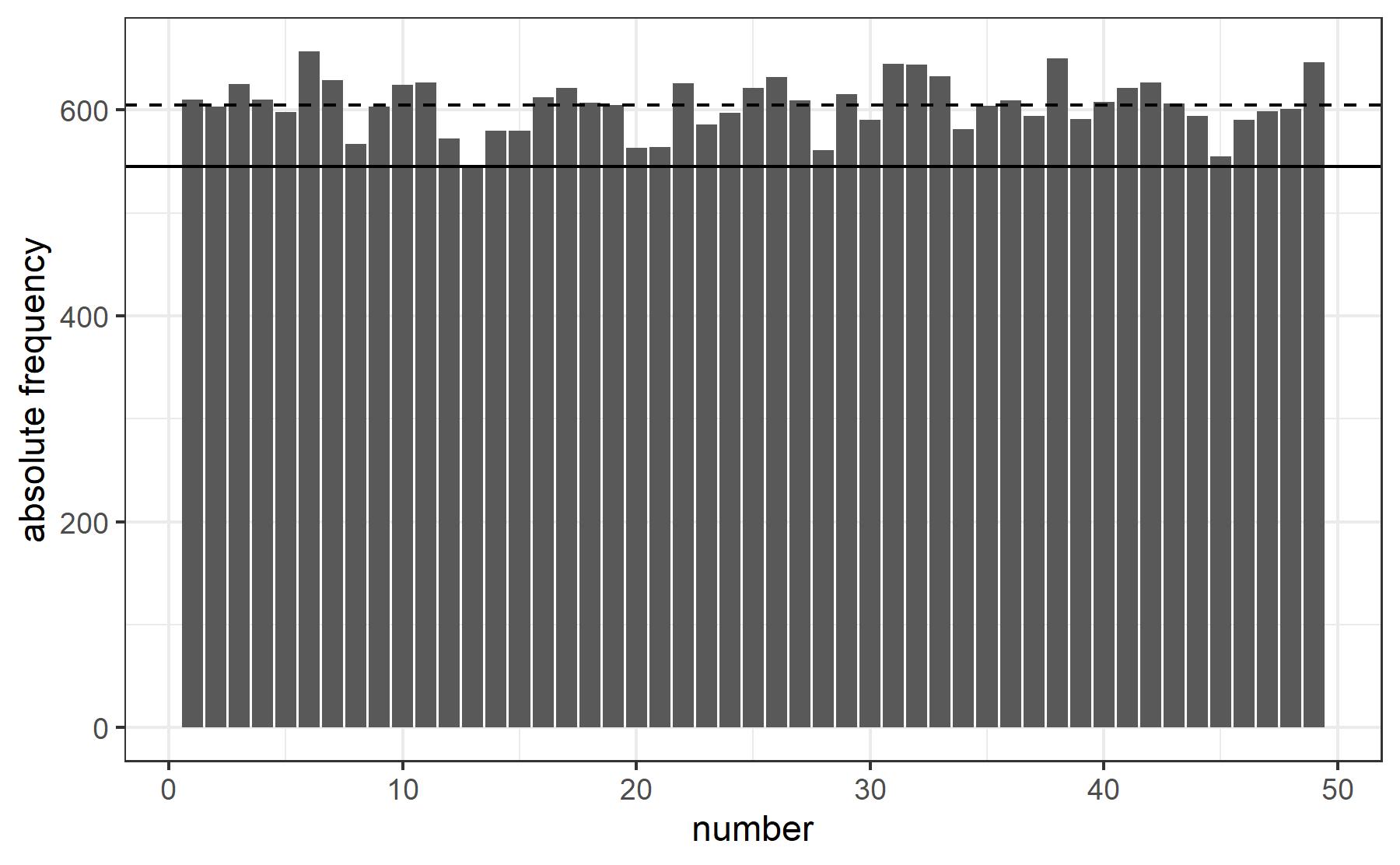}
	\caption{Frequency distribution for $N=4337$ games, where 3615 out of 4337 games included a 7th additional number, leading to $n=29637$ number}
	\label{fig2}	
\end{figure}

The firm operating the lottery may not be happy with a $p$-value of 9.11\%, which is below the 10\% hurdle that many researchers maintain for a semblance of significance. To leave no doubt about the  uniform distribution, the Lotto operator might wish to produce weaker evidence, that is a larger $p$-value. This can easily be achieved by modifying the sample. Notwithstanding the name ``6 out of 49'', in 3615 out of 4337 games an additional number was drawn, which changed the price money,\footnote{On June 17, 1956, the drawing of the 7th additional number was introduced, and abolished on May 4, 2013.} amounting in fact to 7 numbers drawn from 49 without replacement. In Figure \ref{fig2} we display the frequency distribution of this modified sample, which we call sample II in contrast to the original sample, which we call sample I, see Table \ref{tab_lotto} for an overview. Employing equation (\ref{chi_mod_K}) again, we calculate $\chi^2_{(6)}= 59.17$ and $\chi^2_{(7)}= 60.58$ with $p$-values 0.1295 and 0.1051, respectively. Note that neither $\chi^2_{(6)}$ nor $\chi^2_{(7)}$ are fully appropriate for the data behind Figure \ref{fig2}, where $K$ was equal to 7 only in about 83\% of the games and $K$ was 6 else. The correct $p$-value will hence be between 10.5\% and 12.9\%, but above 10\% in any case, jumping over the significance threshold and reducing the evidence against a violation of uniformity.

\begin{figure}
	\noindent
	\centering{}\includegraphics[scale=0.6]{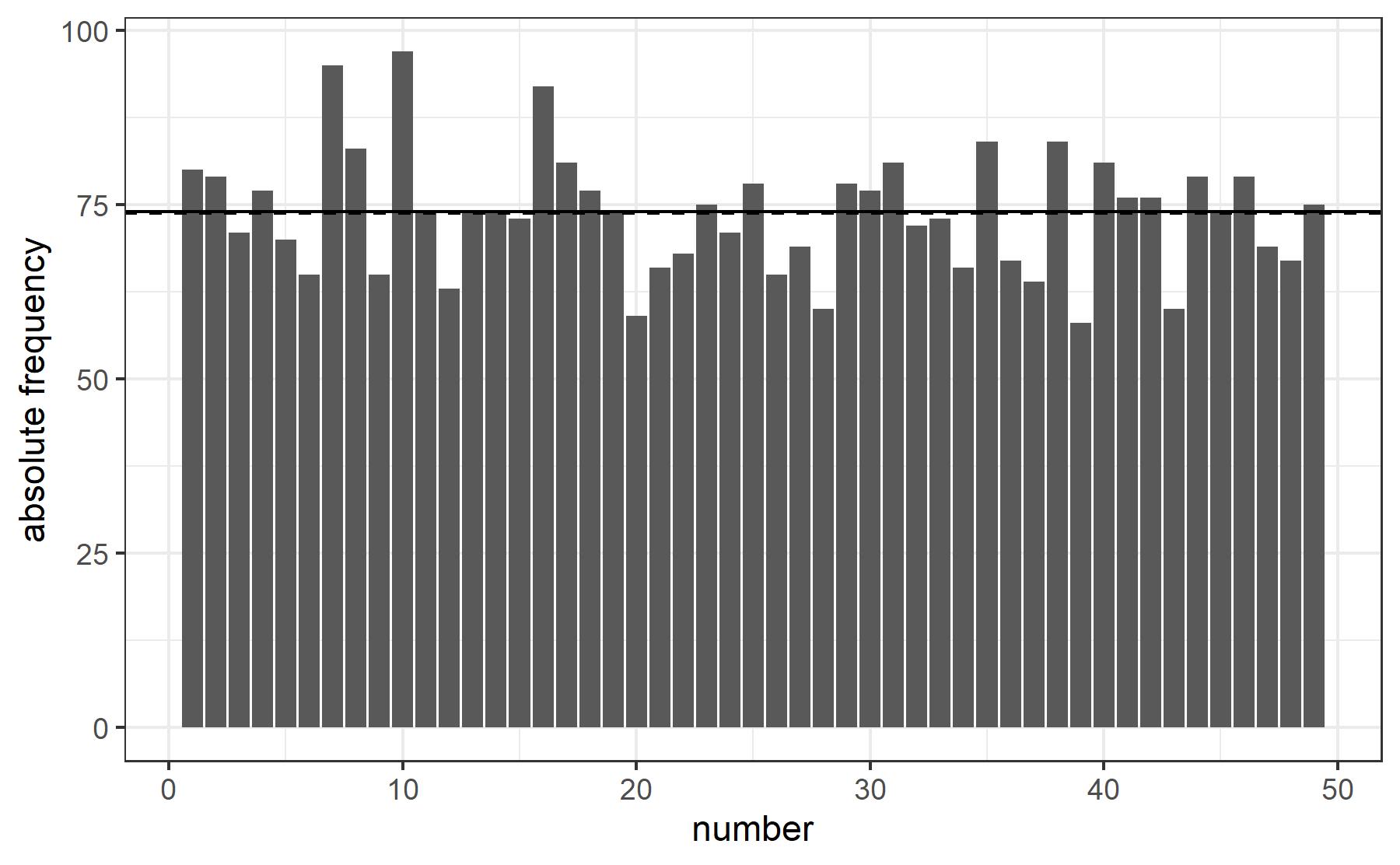}
	\caption{Frequency distribution of the additional numbers from 3615 games}
	\label{fig3}	
\end{figure}

If the rationale is really to minimize the evidence against uniformity, i.e.\ to be as far away from significance as possible in terms of the $p$-value, the lottery firm may come up with some arguments to restrict the attention to the sample of additional numbers only, which we call sample III. This might be seemingly justified by the fact that this sample of size $n=3615$ is independent such that the conventional $\chi^2_{iid}$ from (\ref{chi_iid}) may be used for testing. With the data behind Figure \ref{fig3} this yields $\chi^2_{iid} = 48.64$ with a $p$-value of 44.7\% being beyond any reasonable significance level. 

The strong significance of $Z_{13}^{(6)}$ computed from sample I (and the dramatic change in the distribution of the tests statistic), the much weaker significance of $\chi^2_{(6)}$  and $\chi^2_{(7)}$ from sample I and sample II, respectively, and the insignificance of $\chi^2_{iid}$ from sample III illustrate the effects of MESSing and how MESSies may proceed in practice. One way occurring here is looking at the sample and manipulating the hypothesis or choice of test, which happens with $Z_{13}^{(6)}$; another way consists of extending (or reducing) the sample with seemingly reasonable arguments until $p$-values become significant or insignificant, depending on the goals of the researcher, which happens when moving from sample I to sample II. When moving to sample III or also when picking $m=13$ for $Z_{m}^{(6)}$, the goal is not merely to jump over some (in)significance threshold, but to drive evidence in terms of $p$-values to the extremes, i.e.\ in the one case minimizing and in the other maximizing evidence. The next section provides a broader discussion of several forms of MESSing and connects it to different strands of the literature.

\section{Manipulating Evidence Subject to Snooping}

By coining the term we want to stress that it is crucial to distinguish MESSing from exploratory data analysis and data mining on the one hand and outright fraud on the other hand. The term ``data mining'' has undergone a considerable change in meaning. \citet{Lovell1983}  used it to describe the process of ``experimentation'' until $t$-statistics turn significant, sometimes called data grubbing or dredging or fishing to defame the applied practice of others, see also \citet[p. 1098]{White2000}: ``Although data mining has
recently acquired positive connotations as a means of extracting valuable relationships
from masses of data, the negative connotations arising from the ease
with which naive practitioners may mistake the spurious for the substantive are
more familiar to econometricians and statisticians.'' However, nowadays, data mining receives a lot of attention as a smart toolkit for computational data analysis intersecting with machine learning and engineering techniques like artificial intelligence in order to unveil hidden association or patterns in big data sets, see e.g.\ \citet{Hastieetal2009} for an appreciation. Thus, the term data mining is today often interpreted as a form of exploratory data analysis for big data (see \cite{Hand1998}). Exploratory data analysis is crucial for virtually any statistical analysis and in other applications the data set is often not large enough to split the sample for exploratory analysis and inference, thus data snooping is ``endemic'' not only in time series analysis (see \cite{White2000}). MESSing is the dark side of the perfectly sensible and necessary practices of exploratory data analysis, data mining or snooping; it describes when they may become harmful, namely when they are not acknowledged. On the other hand, even though not acknowledging data snooping is clearly a manipulation, it should be distinguished from outright fraud like e.g.\ fabrication of data. Often the MESSies will not have bad intentions, but e.g.\ simply strive to find evidence in favour of a theory they believe in or follow standard practices in their field. It seems not to be useful to incriminate MESSies, but to convince them to refrain from MESSing. 

We now discuss some forms of MESSing and related issues in detail: One way to MESS is HARKing (Hypothesizing After the Results are Known) defined by \citet[p. 197]{Kerr1998} as ``[...] presenting post hoc hypotheses in a research
report as if they were, in fact, a priori hypotheses''. Such a behaviour is also often called the Texas sharpshooter fallacy, where the shooter paints the target after  firing a shot. Generally, statistical tests are invalidated when one postulates hypotheses or test statistics subject to  data snooping and uses the same data to test them, since ``agreement between a sample and a hypothesis
based on that sample is purely tautological and proves nothing but accuracy in
reading and restating the data of the sample.'', see \citet[p. 229]{Wallis1942}. The traditional approach of statistical testing had been designed for what \citet[p. 112]{Hand1998} called primary data analysis: ``[...] the data are collected with a particular question or set of
questions in mind''. This is the reason why subfields like sampling theory and experimental or survey design are central to statistical theory and practice. \citet{Hand1998} distinguished secondary data analysis defined as ``[...] the process of secondary analysis of
large databases aimed at finding unsuspected relationships
which are of interest or value to the database owners''. \citet{Kerr1998} and \citet{Rubin2017} demonstrated that data mining, or secondary data analysis, invalidates hypothesis testing: A hypothesis that has been postulated only after explorative data inspection must not be tested with the same data. This is why \citet{HollenbeckWright2017} distinguished between THARKing (transparently HARKing) and SHARKing (secretly HARKing).  They classified  SHARKing  as an  unethical practice. Clearly, not all researchers share this view, and SHARKing may be a widespread ``questionable research practice'' as investigated by \citet{Johnetal2012} in an anonymous survey of more than 2.000 academic psychologists. \citet[Table 1]{Johnetal2012} observed that 35\% affirmed  of ``reporting an unexpected
finding as having been
predicted from the start'', which is in the spirit of SHARKing of course.

A further popular form of MESSing is what has been called  $p$-hacking recently, see  \citet[p. 534]{Simonsohnetal2014}: dredging the data until the $p$-value is small enough to reject; see also \citet{Simmonsetal2011}. That way one may produce ``spectacular results'' and catch attention of a wider public. But also in the smaller scientific community, there are strong incentives to produce ``false positive'' results, due to the so-called publication bias. \citet[p. 30]{Sterling1959} already stated that: ``There is some evidence that in fields where statistical tests of significance are commonly used, research which yields nonsignificant results is not published. Such research being unknown to other investigators may be repeated independently until eventually by chance a significant result occurs - an `error of the first kind' - and is published.''; see also \citet{Sterlingetal1995}. In times where researchers are under increasing pressure to publish successfully, HARKing and $p$-hacking may be all the more tempting since ``negative results'' (nonrejection of hypotheses) are hard to publish.

Sometimes, however, the opposite may be true as well: A researcher may be happy not to reject a null hypothesis. Consider specification testing of certain assumptions behind a model we wish to apply. If e.g.\ we want to perform a simple analysis of variance (ANOVA), the underlying assumptions are normality of the data and variance homogeneity. A conscientious statistician would check these assumptions before applying the ANOVA $F$-test, and he or she might be tempted to weaken evidence against the underlying assumptions to jump over the chosen significance level. Such a behaviour has been called reverse $p$-hacking by \citet{Chuardetal2019}. Reverse $p$-hacking might also be observed in research influenced by industries, which are interested in weakening evidence of negative effects of their products on health, e.g.\ of cigarette smoking on lung cancer (see e.g.\ \cite{white2010}).

There may be several reasons why researchers do not only want to reach $p$-values just below (or above) classical significance levels as in the case of (reverse-)$p$-hacking, but to really minimize (or maximize) $p$-values. For example smaller $p$-values often are perceived as lending more credibility and importance to results without even looking at the respective effect sizes (see e.g.\ \cite{WassersteinLazar2018}) or being far away from significance levels could lead to being above suspicion of $p$-hacking. Practices trying to minimize $p$-values have been called ambitious $p$-hacking by \cite{simonsohn2015}. 

Let us now revisit the example of the German lottery from the previous section: A THARKer looks  at Figure \ref{fig1} first, finds the deviations of e.g. $S_{13}$ and $S_6$ ($S_6 = \max_{m=1, \ldots, 49} S_m$) striking and  postulates $H_0$: $p_{m} = 1/49$ for all $m$ in order to see how strongly the data violate the null of uniformity; but then he or she uses  $\chi^2_{(6)}$ as an appropriate test statistic, and there is nothing wrong with this approach, even though snooping took place. A SHARKer, however, would mean to hide the secret look at Figure \ref{fig1}, present the pair $H_0$: $p_{13} = 1/49$ versus $H_1$: $p_{13}< 1/49$ as if it were generated from the prior theory of unlucky numbers, and then use $Z_{13}^{(6)}$ in order to produce a highly significant result. At the same time, this SHARKer maximizes evidence against uniformity in order to report the most spectacular violation. Moving from sample I to sample II, or from $\chi^2_{(6)}$ to $\chi^2_{(7)}$, is in the spirit of reverse $p$-hacking, see Table \ref{tab_lotto}. Restricting the data to sample III, however, is in the spirit of minimizing evidence against the null hypothesis, see again the maximum $p$-value in Table \ref{tab_lotto}. Except for THARKing, all these strategies are forms of MESSing.

To avoid the negative effects of SHARKing or $p$-hacking or of MESSing in general, \citet[p. 132]{WassersteinLazar2018} demanded more transparency from scientific authors: ``Researchers should disclose the number of hypotheses
explored during the study, all data collection decisions,
all statistical analyses conducted, and all $p$-values computed.'' On top of requirements for authors, \citet[p. 1363]{Simmonsetal2011} added guidelines for reviewers and recommended not to push the authors to provide highly significant results. Similarly, \citet[p. 111]{Sterlingetal1995} encouraged journal editors to accept or reject empirical studies irrespective of their outcomes and to make decision rather in light of the importance of the research question and the adequacy of the employed methods and data. In 2015, the editors of \emph{Basic and Applied Social Psychology} (BASP) went one step further and virtually banned ``$p$-values, $t$-values, $F$-values, statements  about  `significant'  differences  or  lack thereof, and so on'' from this journal, see the Editorial by \citet[p. 1]{TrafimowMarks2015}. Unfortunately, a ban of signifiance rituals opens different routes to questionable research practices. In particular, \citet[p. 374]{Frickeretal2019} found when assessing all papers published in BASP in 2016 ``[...] multiple instances of authors overstating conclusions beyond what the data would support if statistical significance had been considered.'' Other measures against MESSing are e.g.\ using methods that are robust to certain forms of it (see e.g.\ \cite{White2000} or the Pearson test in our Lotto example) or preregistration of research designs (see e.g.\ \cite{christensen2019}). Using Bayesian methods is no remedy against MESSing (see e.g.\ \cite{simonsohn2014}).

\section{Small Manipulations, Big Effects}

We propose a simple theoretical example that illustrates MESSing in both directions, i.e.\ strengthening as well as weakening evidence subject to snooping and that makes clear that one is in a sense the mirror image of the other. We then use it to show how large the detrimental effects of these practices can be in terms of size and power distortions even in this simplistic setup.

Consider three researchers interested in assessing Gaussianity of daily stock returns. The first researcher hopes to find a significant deviation from Gaussianity to increase chances for publication of his article on non-normal return distributions. The second researcher is funded by the financial industry, which has no interest in higher capital requirements due to the increased risk implied by e.g.\ fat tails of the return distribution, and thus is looking for insignificant results. The third researcher is just interested in scientific progress.  All researchers use the same data set with $n$ daily stock returns $x_1, x_2, ..., x_n$.

We assume for simplicity that the returns are independent and consider three common and closely related tests of the null of Gaussianity based on the skewness of normal random variables being 0, $\gamma_1=0$, and the kurtosis being 3, $\gamma_2=3$. The tests essentially assess the deviations of the empirical analogues of these standardized moments, $\hat{\gamma}_1$ and $\hat{\gamma}_2$, from 0 and 3, where
\[
\hat \gamma_k = \frac{\frac{1}{n} \sum_{i=1}^n (x_i - \overline x)^{k+2}}{d^{k+2}},\  k=1,2 \quad \mbox{with } d =\sqrt{\frac{1}{n} \sum_{i=1}^n (x_i - \overline x)^{2}} \, .
\]
The first two tests use only one of the two standardized moments, the third test, the Jarque-Bera test, see \cite{jarque1980}, uses both.  The skewness and kurtosis test, see \cite{shapiro1968}, use the squares,
\[
\Gamma_1^2 = n \frac{\hat \gamma_1^2}{{6}} \quad \text{and}  \quad \Gamma_2^2 = {n} \frac{(\hat \gamma_2 -3)^2}{{24}} \, ,
\]
respectively, as test statistics, which asymptotically follow a $\chi^2 (1)$-distribution under the null and the Jarque-Bera test uses the sum of the squares,
$$JB= \Gamma_1^2 + \Gamma_2^2,$$
which follows a $\chi^2 (2)$-distribution asymptotically under Gaussianity.

The third researcher uses the full information, i.e.\ the $JB$ statistic, to test at level $\alpha$ and thus rejects the null if $JB > \chi^2_{1-\alpha}(2)$. The first researcher, striving to maximize evidence, checks first and secretly which sample moment violates the null most and then picks the skewness or kurtosis test accordingly, i.e.\ determines $\Gamma_{max}^2 := \max_{k=1,2} (\Gamma_k^2)$, and rejects if $\Gamma_{max}^2 > \chi^2_{1-\alpha}(1)$. The second researcher, striving to minimize evidence, checks first which sample moment violates the null least and then picks the test in his favour, i.e.\ determines $\Gamma_{min}^2 := \min_{k=1,2} (\Gamma_k^2)$, and rejects if $\Gamma_{min}^2 > \chi^2_{1-\alpha}(1)$.

The MESSing executed by the first and second researcher distorts the properties of the tests, which we quantify now. To calculate the sizes we assume for simplicity that we have a rather large sample such that all the underlying three tests have a real size virtually equal to the nominal size $\alpha$. Under Gaussianity, $\Gamma_1$ and $\Gamma_2$ are independent. Let $R^{(k)}$ be the event that test $k$, $k=1,2$, yields a significant result, i.e.\ $\p(R^{(k)}) = \p(\Gamma_k^2 > \chi_{1-\alpha}^2(1)) = \alpha$. It then holds:
\[
\p (JB > \chi^2_{1-\alpha}(2)) =  \alpha \, ,
\]
\begin{eqnarray*}
	\p (\Gamma_{max}^2 > \chi_{1-\alpha}^2(1)) &=& \p(R^{(1)} \cup R^{(2)})\\
	&=& \p(R^{(1)}) + \p(R^{(2)}) - \p(R^{(1)}) \p(R^{(2)})\\
	&=& 2 \alpha - \alpha^2 ,
\end{eqnarray*}
\begin{eqnarray*}
	\p (\Gamma_{min}^2 > \chi_{1-\alpha}^2(1)) &=& \p(R^{(1)} \cap R^{(2)})\\
	&=&  \p(R^{(1)}) \p(R^{(2)})=  \alpha^2 .
\end{eqnarray*}
For the nominal size of $\alpha=5\%$, the actual sizes of $\Gamma_{max}^2$ and $\Gamma_{min}^2$ are 9.75\% and 0.25\%, respectively. Thus, MESSing through choosing a test in favour of the researchers' targets can almost double the size or let it almost disappear, and weakening evidence subject to snooping is as harmful as strengthening.

To illustrate the change in power, we assume that the null is wrong in the direction of a fat-tailed alternative, namely a $t$-distribution with 10 degrees of freedom, a $t(10)$-distribution. We use a significance level of $\alpha = 0.05$ and obtain the rejection probabilities using simulations with sample size $n=10^3$ and $10^6$ replications. The $t(10)$-distribution is symmetric and leptokurtic ($\gamma_1 =0$ and $\gamma_2=4$). As the deviation from normality is with respect to the fourth moment only, the kurtosis test has a very high power and the skewness test has a low power,
$$\p(R^{(2)}) = 0.9560 \ \mbox{ and } \p(R^{(1)}) = 0.2782,$$
while the JB-test has a power of
$$\p (JB > \chi^2_{1-\alpha}(2)) = 0.9405.$$
Consequently, weakening evidence by essentially picking the skewness test is very effective here,
$$\p (\Gamma_{min}^2 > \chi_{1-\alpha}^2(1)) = 0.2747.$$
Strengthening evidence leads to a slight increase of the already high power of the JB-test used by the honest researcher,
$$\p (\Gamma_{max}^2 > \chi_{1-\alpha}^2(1)) = 0.9594.$$ 
The simulated sizes and powers for $\alpha=0.05$ are summarized in Table \ref{JB_table}.	

\begin{table}[]
	\centering
	\begin{tabular}{@{}llllll@{}}
		\toprule
		test statistic	& $\Gamma_1^2$ & $\Gamma_2^2$ & $JB$ & $\Gamma_{max}^2$ & $\Gamma_{min}^2$ \\ 
		\midrule
		size	&  0.0495 & 0.0460 & 0.0485 & 0.0910 & 0.0046 \\
		power	& 0.2782 & 0.9560 & 0.9405 & 0.9594 & 0.2747  \\ \bottomrule
	\end{tabular}
	\caption{Simulated sizes and powers of the normality tests for a level of $\alpha=0.05$ and a $t(10)$-distribution under the alternative}
	\label{JB_table}
\end{table}

In our example, we allow for only one researcher degree of freedom, namely the choice of the test statistic. Still in this simple case, Table \ref{JB_table} demonstrates that MESSing in both directions may already have drastic consequences. As researchers usually have many degrees of freedom (see \cite{Simmonsetal2011}), the example hints at how serious the consequences of MESSing may be.

\section{Concluding Remarks}

Data mining is a useful and essential tool to cope with the challenges of growing capacities to store and process massive amounts of data. This is not what our note is about. We rather wish to stress a potential downside of data snooping in connection with statistical inference. Our empirical exercise with data from the German lottery and the theoretical example on testing for normality reinforce how misleading statistical hypothesis testing subsequent to  data snooping can be and in how many different forms MESSes (manipulations of evidence subject to snooping) may come along. 


\bibliography{literature_unlucky13}


\section*{Appendix}

The problem of the Bernoulli variates $X_{m,i}$ not being independent and consequently their sum $S_m$ not being binomially distributed and their standardized sum $Z_m^{iid}$ not being asymptotically standard normal can be cured by not considering the single balls drawn. Instead, define for the $j$th game the Bernoulli random variables  indicating if the number $m$ shows up in this game:
$$Y_{m,j} = \sum_{k=K(j-1)+1}^{Kj} X_{m,k} , \quad j=1, \ldots, N .$$
These variables are independent Bernoulli variates, 
$Y_{m,j} \sim Be \left( \frac K {49} \right)$, and by construction they determine the total counts:
$$S_{m} = \sum_{j=1}^N Y_{m,j}.$$ 
Therefore, $S_m$ obeys the following binomial distribution:
$S_{m} \sim Bi \left(N,\frac K {49} \right)$. 
The binomial  test statistic hence becomes
\begin{equation} 
Z_{m}^{(K)} = \frac{S_{m} - \frac{K \cdot N}{49}}{\sigma_{(K)}} \text{ with } \sigma_{{(K)}}^2 = \frac {KN(49-K)}{49^2},
\end{equation}
which is asymptotically standard normal under the null hypothesis. Obviously, the case of drawing only $K=1$ ball each game yields $Z_{m}^{(1)}= Z_{m}^{iid}$. By comparing the two test statistics  one observes that only the variance changes as formulated in equation (\ref{variances}). Drawing with replacement leads to a negative dependence between the $X_{m,i}$ in one game, which has a balancing effect that explains the drop in the variance compared to the i.i.d.\ case. Thus, the large sample test statistic of the classical binomial test is conservative if applied to this problem and needs to be scaled up as is done in equation (\ref{Zmod}) to retain limiting standard normality.

\end{document}